\documentclass[letter,preprint,aps,showpacs]{revtex4}

\usepackage{graphics}
\usepackage{graphicx}
\usepackage{amssymb}
\usepackage{amsmath}
\usepackage{epsfig}
\usepackage{color}
\usepackage{ulem}
\usepackage{hhline}
\usepackage{multirow}
\usepackage{epstopdf}

\begin{document}

\title{Ambient Pressure Superconductivity emerging in the Local Moment Antiferromagnetic Phase of Ce$_{3}$PdIn$_{11}$.}

\author{
M.\ Kratochv\'{\i}lov\'{a}$^{1}$,
J.\ Prokle\v{s}ka$^{1}$,
K.\ Uhl\'{\i}\v{r}ov\'{a}$^{1}$,
M. Du\v{s}ek$^{2}$,
V.\ Sechovsk\'{y}$^{1}$, and
J.\ Custers$^{1}$
}

\affiliation{$^1$ Charles University in Prague, Faculty of Mathematics and Physics, Ke Karlovu 5, 121 16 Prague 2, Czech
Republic.}
\affiliation{$^2$ Department of Structure Analysis, Institute of Physics ASCR, v.v.i., Na Slovance 2, 182 21  Prague, Czech Republic}


\begin{abstract}
We report on magnetization, specific heat and resistivity experiments on single crystals of the novel heavy Fermion compound Ce$_3$PdIn$_{11}$. At ambient pressure the compound exhibits two successive transitions at $T_{1} = 1.63$~K and $T_{\mathrm {N}} = 1.49$~K into incommensurate and commensurate local moment antiferromagnetic states, respectively, before becoming superconducting below $T_{\mathrm {c}} = 0.42$~K. The large values of $dB_{\mathrm {c2}}/dT$ and $B_{\mathrm {c2}} = 2.8$~T imply that heavy quasiparticles form the Cooper pairs. Thus, Ce$_{3}$PdIn$_{11}$ is the first ambient pressure heavy Fermion superconductor where $4{\it f}$ electrons are simultaneously responsible for magnetic order and superconductivity.
\end{abstract}
\date{\today}

\maketitle

Superconductivity in the copper oxides, iron pnictides and heavy fermion (HF) materials emerges from the proximity to a zero temperature magnetic transition, so called quantum critical point (QCP). The shared debate in these disparate classes of materials concerning the pairing mechanism is the nature of the magnetism and its relationship to superconductivity. In terms of clarifying this debate, HF materials have proven to be a fruitful playground~\cite{Loehneysen2007,Pfleiderer2009}. It was demonstrated that in these systems magnetism can evolve either through a spin-density wave (SDW) induced by Fermi surface nesting of itinerant {\it f}--electrons, and consequently the SC in a way that is analogue to that of conventional superconductors with spin fluctuations taking over the role of phonons (e.\,g.\,, Ref.~\cite{Monthoux2001}), or the magnetism has a different origin namely ordering of local moments. This has dramatic consequences. At the QCP the electronic quasiparticles disintegrate leading into a sudden reconfiguration of the Fermi surface. The concepts of ``nesting" and ``gluing" become meaningless requiring a whole new description of electron pairing. More intriguing, in CeRhIn$_{5}$ superconductivity and local moment (LM) magnetism coexist in a certain region of the phase diagram~\cite{Shishido2005,Park2006} implying that a single 4{\it f} state has to be both localized, to account for the magnetism, and itinerant due to participation in the superconductivity. Experiments performed on an appropriate HF system that can be probed by a myriad of techniques, in particular spectroscopic ones, could help in deciding between the pairing mechanisms~\cite{Park2008,Flint2010,Yang2012,Bodensiek2013}.


This work reports on single crystals of Ce$_{3}$PdIn$_{11}$ (or simply 3111). We will show that in this stoichiometric HF compound, local moment antiferromagnetism and superconductivity (SC) compete at ambient pressure. The compound which has been synthesized in polycrystalline form together with Ce$_{5}$Pd$_{2}$In$_{19}$ only recently~\cite{Tursina2013}, is a new member in the Ce$_{n}T_{m}$In$_{3n+2m}$ family. It can be regarded as ``intermediate step" between the cubic CeIn$_3$ and tetragonal Ce$_{2}$PdIn$_{8}$. Ce$_{3}$PdIn$_{11}$ crystallizes in the typical tetragonal structure (space group $P4/mmm$) based on the AuPt$_{3}$-type (CeIn$_{3}$-block) and PtHg$_2$-type ($T$In$_{2}$) units alternating along the $c$-axis as depicted in the upper inset of Fig. 1. Plate-like single crystals with mass up to $\sim$ 1~mg were grown by indium self-flux method using high purity starting materials (Ce 99.9\% and additionally purified by SSE, Pd 99.995\% and In 99.999\%) in the ratio 3:1:25-50 (for details concerning growth and sample homogeneity see Ref.~\cite{Marie2014}). From single crystal X-ray diffraction data refinement we obtained the lattice parameters yielding $a = 4.6896(11)$~\r{A} and $c = 16.891(3)$~\r{A} in good agreement with literature data~\cite{Tursina2013}. A remarkable peculiarity of the structure is that it possesses two inequivalent crystallographic Ce-sites. The Ce2-site (Wyckoff notation: 1$a$) exhibits the proper CeIn$_{3}$ environment, whereas the Ce1-site occupies the 2$g$ position. Its surrounding atoms are identical with the Ce-atoms in Ce$_2$PdIn$_8$. Only the distances are slightly increased. The synthesis of the respective non-magnetic La and Y--3111's was unsuccessful so far. Experiments were performed on samples from different batches.\\

Figure~1 plots the inverse magnetic susceptibility $\chi^{-1}=H/M$ as a function of temperature for a field of 1~T along different crystallographic directions. Measurements show only little anisotropy in the susceptibility between [100] (triangles in Fig.~1) and [001] (bullets) which points to the [001] direction (i.\,e. $c$--axis) being the easy axis of magnetization. No significant anisotropy is found in the (001) plane. The almost isotropic behavior suggests a strong influence of the cubic CeIn$_{3}$ unit on magnetic correlations in this compound. In line with other members of this family of HF materials, CeIn$_{3}$, CeRhIn$_{5}$ and Ce$_{2}$RhIn$_{8}$, it can be assumed that the in-plane nearest neighbor Ce--moments couple antiferromagnetically~\cite{Benoit1980,Lawrence1980,Bao2000,Bao2001}. At temperatures above $T \approx 85$~K for $B \parallel a$ and above $T \approx 100$~K for $B \parallel c$, $\chi(T)$ follows a Curie-Weiss law with effective moment of $\mu_{\mathrm {eff}}= 2.43 \mu_{\mathrm {B}}$ for both directions. The experimental values of $\mu_{\mathrm {eff}}$ are slightly reduced from Hund's rule value of $2.54 \mu_{\mathrm {B}}$ for a Ce$^{3+}$ ion. The respective paramagnetic Weiss temperatures yield $\theta^{a}_{\mathrm{P}} \backsimeq -49$~K and $\theta^{c}_{\mathrm{P}} \backsimeq -33.5$~K. In terms of Kondo-type interactions, neglecting crystal electric field (CEF) effects, the Kondo temperature would be of the order of $T_{\mathrm {K}} \backsimeq |\theta_{\mathrm{p}}|/4 \backsimeq 12$~K~\cite{Hewson1993}. At the lowest temperatures a weak maximum becomes visible in $\chi^{c}(T)$ (see Fig.~1 lower inset). This might be attributed to critical spin fluctuations of a nearby magnetic ordering at lower temperatures.\\
For further characterization of the physical properties of Ce$_{3}$PdIn$_{11}$, we performed specific heat ($C/T$) and resistivity measurements. The results of the high temperature resistivity measurements is summarized in the inset of Fig.~2b. Above 40~K $\rho(T)$ is weakly temperature dependent. It passes than through a local maximum at $T \backsimeq 35$~K before it falls rapidly at lower temperatures to a value of approximately $10 \mu\Omega$cm at $T=2$~K for current $j$ applied along [110] and [001] direction. The residual resistivity ratio $RRR=\rho_{300{\mathrm {K}}}/\rho_{2{\mathrm {K}}}$ equals 5. For $j$ along [100], a slightly higher resistivity is obtained being $16 \mu\Omega$cm at 2~K. Traditionally, this behavior is attributed to the transition from coherent to incoherent Kondo scattering off the Ce-sites and to the influence of excited CEF states of Ce$^{3+}$. Often both contributions are visible in resistivity by the appearance of a double-–maximum structure. The absence of such a characteristic in Ce$_{3}$PdIn$_{11}$ suggests that the CEF and Kondo effect overlap.
The high temperature $C/T$ data (not shown) were fitted in the interval 6~K $<T<$20~K using the simple relation $C/T = \gamma + \beta T^{2}$. Here, $\gamma$ is the Sommerfeld coefficient and $\beta$ is the Debye lattice term. Considering that the $T^2$ approximation to the lattice heat capacity does not account for additional contributions such as spin fluctuations and Schottky-like ones, we realize the limitation of the procedure adopted here. The resultant parameters are $\gamma = 290$~mJ/mol-Ce$\cdot$K$^{2}$ and $\Theta_{\mathrm{D}} = 214$~K. The strongly enhanced $\gamma$-value identifies Ce$_{3}$PdIn$_{11}$ as a heavy Fermion compound. The low temperature zero field data are displayed in Fig.~2. Three clear features are visible, marked each by an arrow.  The first anomaly appears at $T_{1} = 1.63$~K going from higher to lower temperature. The value is just below the accessible experimental temperature of our magnetization experiment, but taking the maximum structure in $\chi(T)$ as a guide, $T_{1}$ signals the onset of long-range magnetic ordering. $T_{1}$, however, does not reflect in $\rho(T)$ as evident from Fig.~2b. The first transition is soon followed up by a second one at $T_{\mathrm {N}} = 1.49$~K which results a in shallow decrease in the resistivity. Within the resolution of the experiments neither at $T_{1}$ nor at $T_{\mathrm {N}}$ a hysteresis was detected, suggesting that both transitions are likely of second order. The double peak structure is reminiscent of that in (Ce$_{0.8}$La$_{0.2}$)Ru$_{2}$Si$_{2}$~\cite{Mignot1991} CeRu$_{2}$(Si$_{1-x}$Ge$_{x}$)$_2$~\cite{Besnus1996} and CeRuSiH$_{1.0}$~\cite{Chevalier2008}. These tetragonal compounds exhibit two antiferromagnetic transitions (e.\,g.\, $5.6$~K and $1.8$~K for (Ce$_{0.8}$La$_{0.2}$)Ru$_{2}$Si$_{2}$ and at $7.5$~K and $3.1$~K for CeRuSiH$_{1.0}$). In close analogy we assign the first peak at $T_{1}$ to an incommensurate AFM ordering while the second one at $T_{\mathrm {N}}$ would hence correspond to a transition into a commensurate structure (or locking in).
The most intriguing feature in the specific heat appears around $T_{\mathrm {c}} =0.39$~K and indicates the transition in a superconducting phase as is evident from corresponding resistivity data. The small discrepancy in bulk $T_{\mathrm {c}}$ and the onset of SC in the resistivity ($T_{\mathrm {c}} \approx 0.5$~K) is not unusual, see for example CeIrIn$_{5}$~\cite{Petrovic2001}. In order to accurately determine the superconducting transition we assumed an idealized jump at $T_{\mathrm {c}}$ in accordance with the entropy balance between the normal and superconducting state yielding a $T_{\mathrm {c}} = 0.42$~K (inset in Fig.~2a). Some estimate of the normal state Sommerfeld coefficient $\gamma_{\mathrm {n}}$ may be obtained from a polynomial extrapolation of the specific heat data below $T_{\mathrm {N}}$ intercepting zero at $\gamma_{\mathrm {n}} = 1.52$~J/mol$\cdot$K$^2$. Using these results allows us to calculate the parameter $\Delta C/(\gamma_{\mathrm {n}}T_{\mathrm {c}}) \approx 0.62$. That is roughly half of the expected value from BCS theory ($\Delta C/(\gamma_{\mathrm {n}}T_{\mathrm {c}}) = 1.43$). However, strongly reduced BCS values have been reported for other unconventional superconductors as well, e.\,g.\,, CePt$_{3}$Si and Sr$_{2}$RuO$_{4}$~\cite{Bauer2004,Mackenzie2003}. \\
More insight in the compound's properties comes from the entropy. The magnetic entropy $S(T) = \int_{0}^{T} C_{4f}/T dT$  is shown in Fig.~2a in the upper inset. $C_{4f}$ has been determined by subtracting the Debye contribution from the total specific heat. The entropy gain of about $0.6R\ln2$ at $T=T_{1}$ which transforms to $0.2R\ln2$ per Ce is well below that associated with the lifting of the degeneracy of the ground state doublet and hints to ordering with substantially reduced magnetic moments. More important, the full expected value of $R\ln2$ per Ce-atom is reached at 17~K. This temperature is close to $T_{\mathrm {K}}$ retrieved from susceptibility and is a clear indication of local moment ordering in Ce$_{3}$PdIn$_{11}$.\\
From comparison with for example, CeRuSnH$_{1.0}$~\cite{Chevalier2008} we infer that the magnetic transitions in Ce$_{3}$PdIn$_{11}$ are strongly field dependent. The effect of an applied magnetic field on the $T$-dependence of the specific heat is presented in Figure 3. For $B \parallel c$ and in fields $< 2.5$~T the transition $T_{\mathrm {1}}$ shifts to lower temperatures while $T_{\mathrm {N}}$ remains mainly unaffected (Fig.~3a). At $B \approx 3$~T both transitions seem to merge before they split again in higher fields. Of particular interest is the shape of the lower peak. For $B \geq 5$~T it is sharp compared to low field data, suggesting a possible first order transition. The situation is different when the field is applied perpendicular to $c$-axis. As displayed in Fig.~3b, $T_{1}$ first shows a tiny increase before it continuously decreases with fields $>1$~T. The lower transition $T_{\mathrm {N}}$ is almost constant for $B<5$~T and moves monotonically downwards in larger fields. The evolution of $T_{1}$ and $T_{\mathrm {N}}$ are summarized in Fig.~3c. The $T -B$ phase diagram for $B \parallel c$ has been measured out in more detail by the method of analyzing the response of the compound to thermal heat pulses.
By this method a heat pulse resulting in a $\Delta T = 1$~K temperature change of the sample (mounted quasi-adiabatically) to the bath temperature $T_{\mathrm{bath}}$ was applied. The resulting thermalization of the sample temperature to $T$ was monitored over time $t$. The color plot shows the derivative $d(\Delta T)/dt$ as a function of bath temperature and field (red: small change in $dT/dt$; blue/violet: large change in $dT/dt$).
From Fig.~3c it becomes evident that the two transitions, $T_{1}$ and $T_{\mathrm{N}}$, merge at $\approx 3$~T and separate again in higher fields. In fields $>4$~T the $T_{\mathrm{N}}$ anomaly sharpens as inferred by the rapid change in color code. It supports the observation in $C/T$ (see Fig.~3a). The transition temperatures deduced from specific heat experiments are included in the figure ($T_1$: circles, $T_{\mathrm{N}}$: diamonds).
In essence the $T$--$B$ phase diagram for $B \parallel c$ is similar to the one for CeRuSiH$_{1.0}$~\cite{Chevalier2008}. Here both transitions merge at $B=2.5$~T, and separate again in higher fields. Of particular interest is that in fields $>1$~T the lower transition in CeRuSiH$_{1.0}$ changes from being AFM to ferrimagnetic. Perhaps this also is the origin of the sharp, first-order like anomaly in Ce$_{3}$PdIn$_{11}$ for $B \geq 5$~T.
Anyway, the $B$--dependence of the transitions in Ce$_{3}$PdIn$_{11}$ allows for some speculation about the magnetic structure and supports conclusions drawn from $\chi(T)$ data earlier. The weak response of $T_{1}$ and $T_{\mathrm {N}}$ to field when applied $\perp c$ hints at moments mainly oriented antiferromagnetically within the basal plane. A small tilting of the moments outside the (001)-plane, as a consequence of the Ce(1)--Ce(2) interaction, is interfered by the initial increase of $T_{1}$ which would be due to first aligning the moments in the tetragonal basal plane. This magnetic structure closely resembles that for Ce$_{2}$RhIn$_{8}$~\cite{Bao2001}. It is noteworth that this compound exhibits a field--induced first order phase transition below its antiferromagnetic one~\cite{Cornelius2001}. $T_{\mathrm {N}}$ in Ce$_{3}$PdIn$_{11}$ might stem from (commensurate) coupling of the CeIn$_{3}$ building blocks along $c$-axis. \\
The striking observation of bulk superconductivity emerging in the local moment magnetic phase makes Ce$_{3}$PdIn$_{11}$ unique within the Ce$_{n}T_{m}$In$_{3n+2m}$ family. In other members either pressure was necessary to induce SC within the AFM state, like in CeRhIn$_{5}$~\cite{Pagliuso2001} or vice versa by means of doping an AFM phase was established in a SC material. Examples are CeIrIn$_{5}$ and CeCoIn$_{5}$~\cite{Pham2006}, although as pointed out in a recent report the doping elicited magnetic phase has different origin and hence is not quantum critical~\cite{Seo2013}. To provide additional information about the superconducting phase, we performed resistivity experiments to determine the upper critical field $B_{\mathrm {c2}}$. The field was applied $\perp c$ and the current $j \parallel [100]$. From data shown in the inset of Fig.~4 the superconducting phase diagram for Ce$_{3}$PdIn$_{11}$ was constructed with the midpoint of the resistivity drop with respect to $T_{\mathrm {c}}$. A mean--field type of fit describes the data reasonably well with $B(0)_{\mathrm {c2}} = 2.8$~T and an initial slope of $-dB_{\mathrm {c2}}/dT |_{T = T_{\mathrm {c}}} = 9.6$~T/K. The high value of $B_{\mathrm {c2}}$ and $-dB_{\mathrm {c2}}/dT$ are traditionally taken as additional evidence for HF superconductivity. In comparison however, these values are significantly lower than those determined in other Ce$_{n}T_{m}$In$_{3n+2m}$ members in the vicinity of a magnetic QCP, e.\,g.\,, CeCoIn$_{5}$ ($T_{\mathrm {c}}= 2.3$~K ; $B_{\mathrm {c2}}^{\perp c}= 11.6$~T ; $-dB_{\mathrm {c2}}^{\perp c}/dT = 24$~T/K), CeRhIn$_{5}$ (at $p \approx p_{\mathrm {c}} = 2.45$~GPa: $T_{\mathrm {c}}= 2.2$~K ; $B_{\mathrm {c2}}^{\perp c}= 9.7$~T ; $-dB_{\mathrm {c2}}^{\perp c}/dT = 18.4$~T/K)~\cite{Ida2008} and Ce$_{2}$PdIn$_{8}$ ($T_{\mathrm {c}}= 0.68$~K ; $B_{\mathrm {c2}}^{\perp c}= 4.8$~T ; $-dB_{\mathrm {c2}}^{\perp c}/dT = 14.3$~T/K)~\cite{Kaczorowski2009} which is consistent with the assumption that the 3111 structure is more 3-dimensional and hence $T_{\mathrm {c}}$ is reduced~\cite{Monthoux2001}. On the other hand, the parameters of CeIrIn$_{5}$ are even lower ($T_{\mathrm {c}}= 0.4$~K ; $B_{\mathrm {c2}}^{\perp c}= 0.53$~T ; $-dB_{\mathrm {c2}}^{\perp c}/dT = 2.53$~T/K)~\cite{Movshovich2002} hinting that the distance to the QCP, still to be determined for Ce$_{3}$PdIn$_{11}$, and consequently the strength of magnetic fluctuations, play an important role in the size of $T_{\mathrm {c}}$ as well. Estimating of some important parameters which characterize the unconventional SC state in Ce$_{3}$PdIn$_{11}$ is straightforward. In accordance with common practice using BCS theory and deduced values for $T_{\mathrm {c}}$, $\rho_{2\mathrm {K}}$ and $\gamma$, we find a BCS coherence length of $\xi_{0} \approx 12$~nm as $T \rightarrow 0$. This is comparable to the mean free path of the quasiparticles, $l_{\mathrm {tr}} = 45$~nm. The London penetration depth yields a rather high value, i.\,e.\,, $\lambda_{\mathrm {L}} = 922$~nm (for $T \rightarrow 0$). The Ginzburg-Landau parameter in the "pure limit" ($l \gg \xi_0$) is estimated to be $\kappa_{\mathrm {GL}} \approx 82$. Such a high value is common for high-$T_{\mathrm {c}}$ but found in HF materials as well (CeCoIn$_{5} = 108$~\cite{Ikeda2001} ; Ce$_{2}$PdIn$_{8} = 21$~\cite{Kaczorowski2009}). \\

In summary, among other 4{\it f} heavy Fermion superconductors, Ce$_{3}$PdIn$_{11}$ occupies a unique position. The compound undergoes two magnetic transitions at $T_{1} = 1.63$~K (incommensurate) and $T_{\mathrm {N}} = 1.49$~K into a commensurate antiferromagnetic state before at lower temperatures a transition into a superconducting phase at $T_{\mathrm {c}} =0.42$~K occurs. Thermodynamic data revealed that the AFM state is induced by local moments resulting from the 4{\it f} electrons which are concurrently forming the heavy quasiparticle Cooper pairs. The emergence of these competing phases at ambient conditions allows the use of novel experimental techniques, e.\,g.\,, STM to investigate their interrelationship and hence holds promise for bridging our understanding of superconductivity in the presence of local magnetic moments. Further in-depth studies are planned to investigate the complex magnetic phase diagram (neutron experiments) and to determine the critical pressures $p_{\mathrm {c1}}(T_{\mathrm {N}} \rightarrow 0)$ and $p_{\mathrm {c2}}(T_{\mathrm {1}} \rightarrow 0)$ and the quantum critical nature of this compound.\\
We would like to thank Philip W. Phillips for valuable suggestions. This work was supported by the Grant Agency of the Czech Science Foundation (Project P203/12/1201). Experiments were performed in MLTL (http://mltl.eu/) which is supported within the program of Czech Research Infrastructures (project no. LM2011025). The crystallography was supported by project P204/11/0809 of the Czech Science Foundation.

\vspace{0.5cm}

\bibliographystyle{prsty}

\begin{figure}[t]
\centerline{\includegraphics[width=\columnwidth]{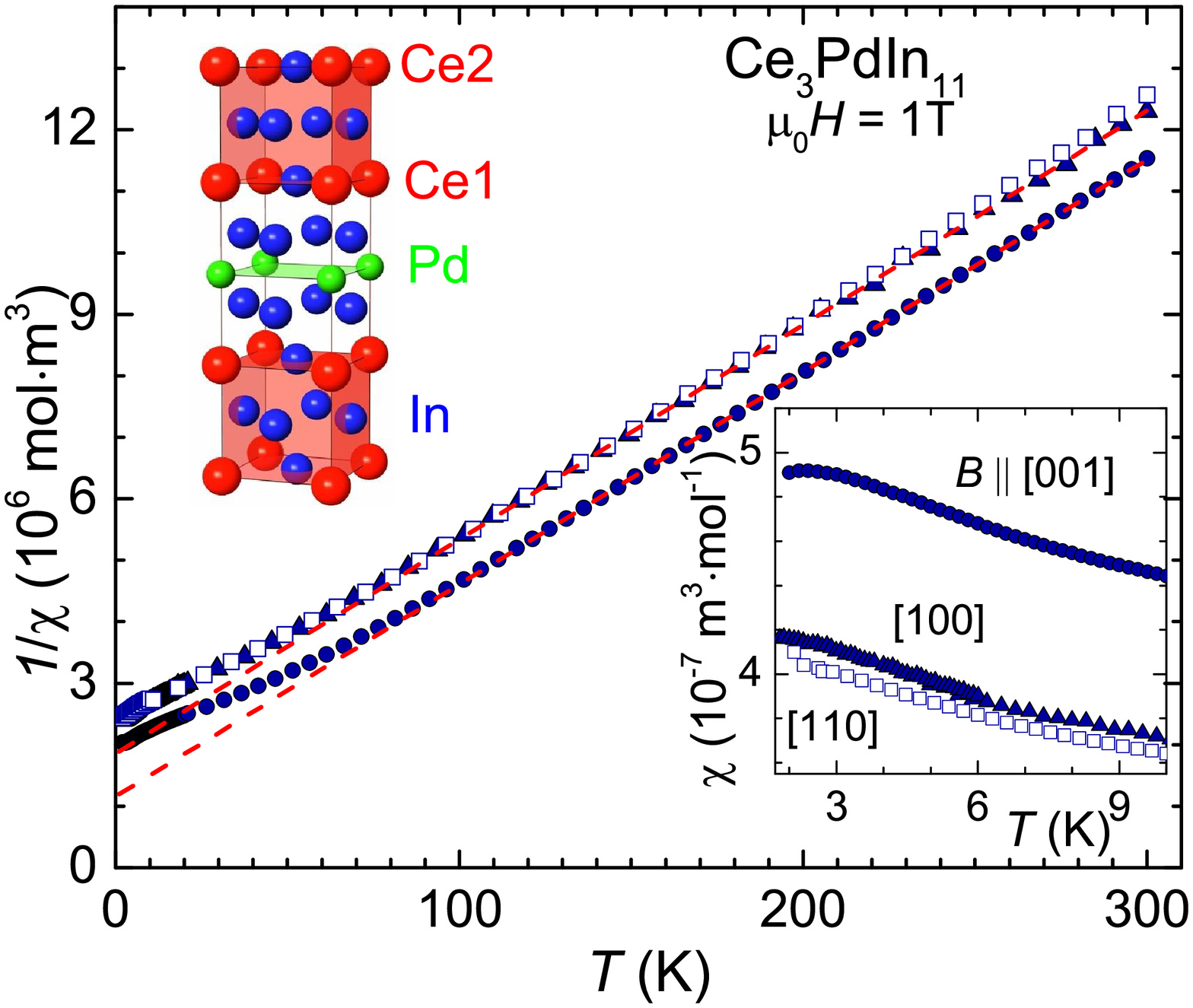}}
\caption{\label{Fig1} Temperature dependence of the inverse susceptibility $\chi^{-1}$. A magnetic field of 1~T was applied $\parallel$ [100] (triangles), $\parallel$ [110] (squares) and $\parallel$ [001] (circles). The (red) dashed line is the Curie-Weiss fit. Upper corner: crystal structure of Ce$_3$PdIn$_{11}$ emphasizing the CeIn$_3$ building blocks. Inset: magnified low-$T$ region $\chi(T)$ in $B=1$~T.}
\end{figure}

\begin{figure}[t]
\centerline{\includegraphics[width=0.8\columnwidth]{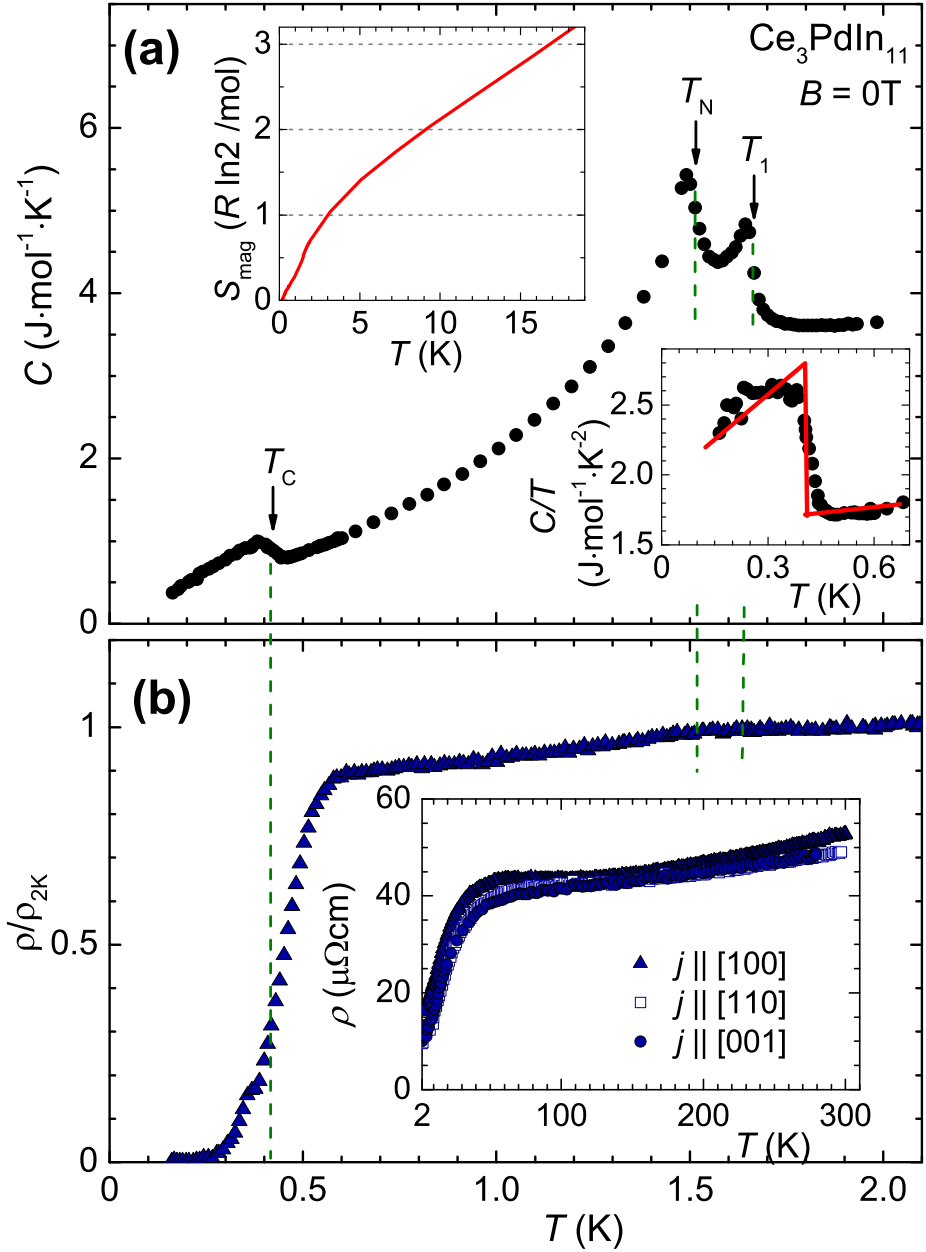}}
\caption{\label{Fig2} (a) The zero field heat capacity $C$ vs. $T$. Upper inset: temperature evolution of the magnetic entropy $S_{\mathrm{mag}}$ in units of $R\ln2$. Lower inset: the equal entropy construction used for the determination of the critical temperature $T_{c}$ for the data taken in $B =0$~T. (b) The temperature dependence of the electrical resistivity normalized to 2~K $(\rho(T)/\rho_{2\mathrm{K}})$ in zero field of Ce$_3$PdIn$_{11}$, measured with $j$ perpendicular to the $c$ axis is shown. Inset: High temperature $\rho(T)$ measured with current applied along the [100] (triangles), [110] (squares) and [001] (circles) in zero field. For this experiment a crystal from a different batch was used.}
\end{figure}

\begin{figure}[t]
\centerline{\includegraphics[width=0.7\columnwidth]{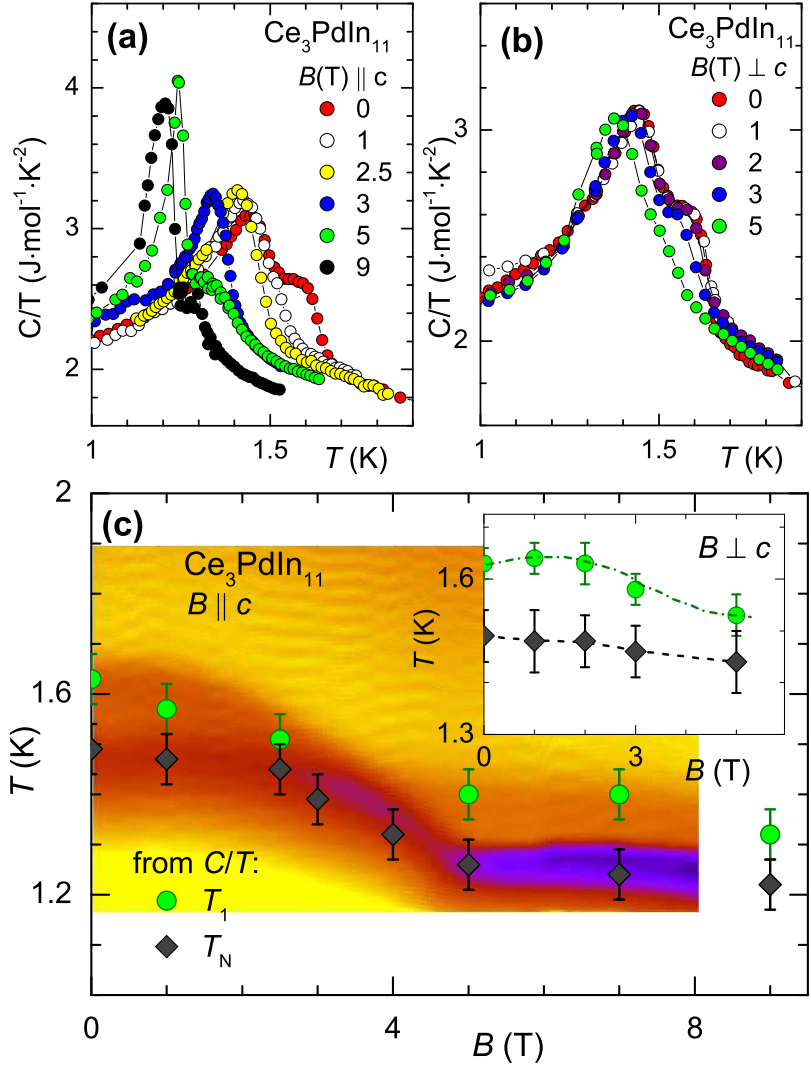}}
\caption{\label{Fig3} (Color online) Specific heat as function of temperature for various values of applied magnetic fields (a) $B \parallel c$ and (b) $\perp c$. The temperature --  magnetic field phase diagram (c) for $B \parallel c$ mapped out by thermal response technique (see text). The (green) circles and (black) diamonds show $T_{1}$ and $T_{\mathrm{N}}$, respectively, obtained in (a). Inset depicts the $T$--$B$ phase diagram for $B \perp c$ axis.}
\end{figure}

\begin{figure}[t]
\centerline{\includegraphics[width=0.8\columnwidth]{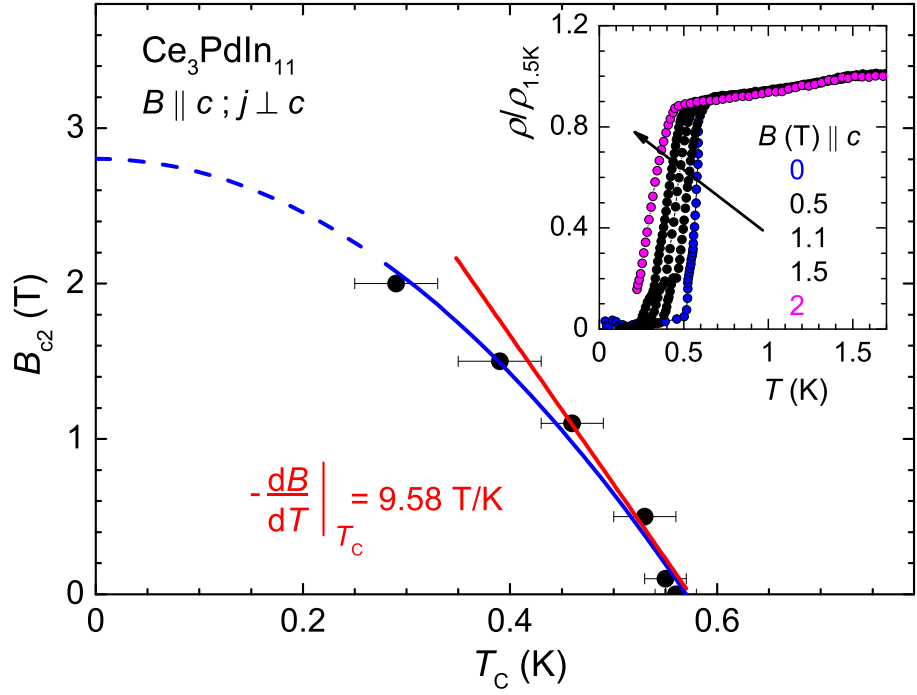}}
\caption{\label{Fig4} Temperature dependence of the upper critical field of $B_{\mathrm{c2}}$ as derived from resistivity experiments on Ce$_3$PdIn$_{11}$. Typical curves are shown in the inset. Current was $j \perp c$ and the field was applied parallel to the $c$-axis. The blue line is a mean-field fit of $B_{\mathrm{c2}}(T)$. The red solid line depicts the slope in the vicinity $B_{\mathrm{c2}} \rightarrow 0$T.}
\end{figure}

\end{document}